# New Thinking on, and with, Data Visualization

Alyssa A. Goodman, Harvard University
Michelle A. Borkin, Northeastern University
Thomas P. Robitaille, Aperio Software Ltd.

**ABSTRACT**

As the complexity and volume of datasets have increased along with the capabilities of modular, open-source, easy-to-implement, visualization tools, scientists' need for, and appreciation of, data visualization has risen too. Until recently, scientists thought of the "*explanatory*" graphics created at a research project's conclusion as "pretty pictures" needed only for journal publication or public outreach. The plots and displays produced *during* a research project--often intended only for experts--were thought of as a separate category, what we here call "exploratory" visualization. In this view, discovery comes from exploratory visualization, and explanatory visualization is just for communication. Our aim in this paper is to spark conversation amongst scientists, computer scientists, outreach professionals, educators, and graphics and perception experts about how to foster flexible data visualization practices that can facilitate discovery and communication at the same time. We present an example of a new finding made using the glue visualization environment to demonstrate how the border between explanatory and exploratory visualization is easily traversed. The linked-view principles as well as the actual code in glue are easily adapted to astronomy, medicine, and geographical information science--all fields where combining, visualizing, and analyzing several high-dimensional datasets yields insight. Whether or not scientists can use such a flexible "undisciplined" environment to its fullest potential without special training remains to be seen. We conclude with suggestions for improving the training of scientists in visualization practices, and of computer scientists in the iterative, non-workflow-like, ways in which modern science is carried out.

## MOTIVATION & BACKGROUND

### A New Conversation
Our aim in this paper is to spark conversation–amongst scientists, computer scientists, outreach professionals, educators, and graphics and perception experts–about how to foster flexible data visualization practices that can facilitate discovery and communication *at the same time*. The paper is loosely based on a talk titled "The Road for Exploration to Explanation, and Back," delivered in March 2018 by Alyssa Goodman at the National Academy of Sciences Sackler Colloquium entitled "Creativity and Collaboration: Revisiting Cybernetic Serendipity." Together, as three authors, we pool our expertise in science, education, data science, visualization research, cognition and software development to offer a mixture of perspectives not commonly found in the literature of any of these areas alone.

### Not just pretty pictures—healing the explanatory-exploratory divide
Most people grasp concepts more quickly from pictures than they do from words. Experienced scientists nearly all read a journal article's abstract, then look at the "pictures" (graphs, charts, images), and then the conclusions before deciding whether or not to read the full article. Readers of technical literature at any level appreciate and learn from images that illuminate the concepts discussed. Recent work on "Viziometrics" demonstrates that more richly illustrated journal articles have higher citation counts (1), and work on memorability shows that figures that have their main point clearly and redundantly encoded in their design make associated concepts more memorable (2). Still, though, many of today's professional scientists think that beautiful graphics are either "only for the public" or not worth the time they take to create. We disagree.

Until relatively recently scientists hand-drew their own sketches of processes, results, and ideas as their experiments, observations, or theorizing progressed. They cleaned up their handmade "exploratory" visualizations for "explanatory" publication, later. Galileo's 1610 drawings of his telescopic observations of the moons of Jupiter offer a famous example of this process, moving from hand-drawn sketches in his diary-like log, to copied-out cleaned-up hand-drawings, to a printed, typeset page (Figure 1).

Over the past century, as science, education, and publishing became increasingly professionalized, specialists were employed to produce clear, well-drawn explanatory graphics to accompany scholarly publications and teaching materials. As a result, a divide started to grow between rough graphics and polished graphics, which facilitated a slow shift of graphics produced as part of "exploratory" data analysis *away* from "explanatory" graphics produced for publishing and public outreach.

Published graphics, especially those intended to have broad reach, are often simplified or altered in ways that make them less accurate than their sketchier predecessors–and this has been a problem since the dawn of publishing. One can determine the orbits of Jupiter's moons more accurately from Galileo's best hand-drawn sketches than from printed pages of *Sidereus Nuncius*, the accuracy of which are limited by typesetting grids (see Figure 1, (3),(4)).

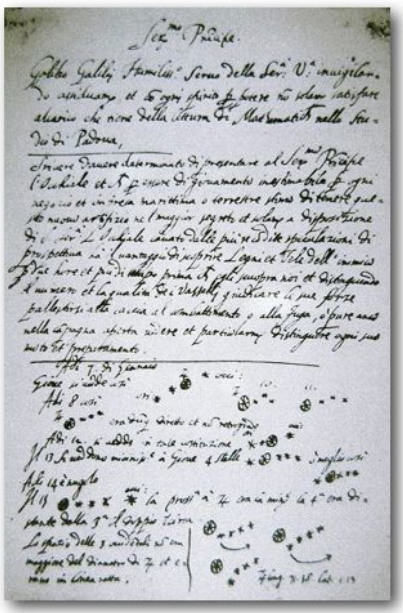 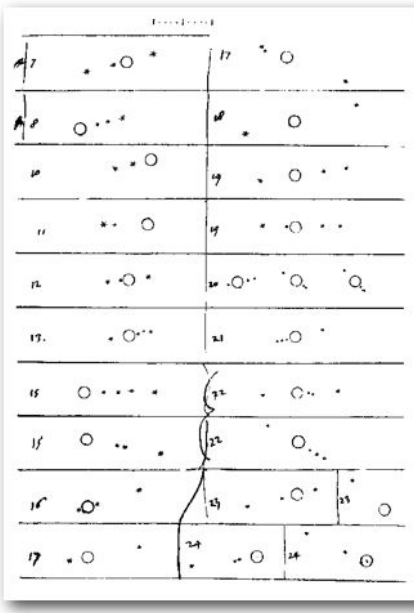 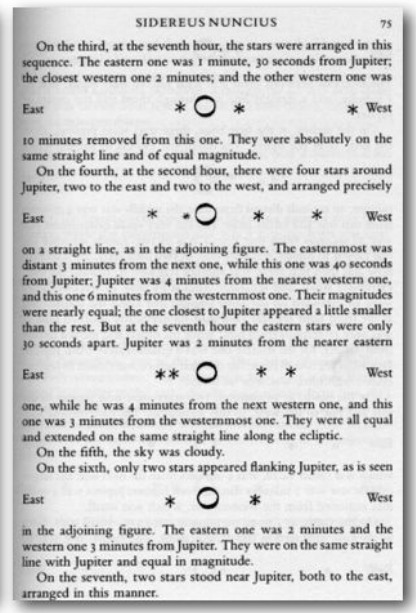

**Figure 1:** Three views of Galileo's Observations of Jupiter's Moons. "Diary" notes, hand-drawn visualization, and printed *Sidereus Nuncius* page (5).

**Visualization Tools as Problem, and Solution**

When tools to produce refined-looking graphics are only accessible to, or usable by, professionals and/or when expert-to-public translation leads to inaccuracies, a fear of elegant-looking graphics, and a concomitant exploratory-explanatory divide is understandable. The divide often leaves behind the aforementioned idea that exceptionally fancy graphics, and the time invested to make them, are only for public consumption, and not really useful for serious scientists or others pursuing deep quantitative analysis.

As this exploratory–explanatory graphics divide widened over the past several decades, so did the size and complexity of the data sets being analyzed by scientists, business analysts, and even students. The start of the modern ramp-up in data scale coincided with an expansion of the so-called "desktop publishing" industry, and by the 1990s most institutions thought software (e.g., Microsoft Excel and PowerPoint) could replace the professionals who used to help make excellent graphics for publication. This unfortunate combination of rapidly-increasing data volume and early do-it-yourself tools, enabled by growing computational capabilities, unfortunately led (in our opinion) to a drop in the quality of journal graphics in many cases. **Access to software and *proficiency* at using it to gain or communicate insight are not the same thing.**

Over the past decade or so, though, researchers and educators in the business of learning and teaching using data have been offered a wealth of new approaches and easily-accessible tools that are reversing the downward trend in graphical communication quality. These new approaches and tools also have the potential to heal the exploratory-explanatory divide. Open source, modular, software is being widely used at many stages of inquiry, and many leaders in technology and science research and education are coming to realize the payoff that training in data visualization[1] can bring, both in terms of new scientific insight, as well as improved communication, both amongst experts, and with the public.

**EASY TRAVEL FROM EXPLORE TO EXPLAIN, AND BACK**

We argue here, as Goodman did in her "Road from Exploration to Explanation, and Back" Sackler talk, that judicious combinations of open-source, modular software tools can create very flexible visualization environments, usable directly by scientists.[2] Thanks to recent dramatic improvements in both software and computing, today's best environments produce visualizations of quality high enough for *both* scientific insight and public outreach. These same software tools, when combined with the increasing availability of data used in published visualizations–whether in scholarly journals or as outreach media–allow the road from exploration to explanation to become a very short two-way highway from "Explore" to "Explain" and back, making it possible even to derive fully new science from "pretty pictures."

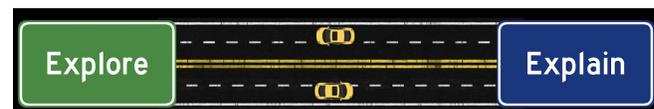

---

[1] This kind of investment in visualization training has been promoted as worthwhile for more than a decade. The 2006 NIH-NSF *Visualization Research Challenges report* (6) put it this way: "Just as knowledge of mathematics and statistics has become indispensable in subjects as diverse as the traditional sciences, economics, security, medicine, sociology, and public policy, so too is visualization becoming indispensable in enabling researchers in other fields."

[2] One of the first instantiations of this open source+modularity principle was the famous "Visualization Tool Kit" (VTK), originally created in 1993, funded by the US Government, improved and maintained by private industry (Kitware), and still in use today (6) (7).

**Why?**
Before we explain how this easy travel between Explore and Explain can happen, let us briefly consider why it might be desirable. There are three reasons, all having to do with time:
1. **"Real-time functionality"** As the statistician John Tukey so elegantly demonstrated and stated when he developed the very first computer-enabled high-dimensional interactive data-visualization system, PRIM-9 (8–10), humans' **spatial recognition abilities work best** when a system responds to user-initiated actions **in real-time.** Data exploration, and user interfaces in general, that are **interactive** and able to live-update changes to the data or view are most effective (11). And, the more attractive and clear a visualization is, the more can be learned from it. So, if "publication-quality" (high-resolution, well-designed) visualizations and images can respond to user actions in real time as part of exploratory tools, then there *is essentially no difference between the product of an exploratory visualization, and one that's suitable for explanation.* In other words, why not use publication-quality graphics in real-time exploratory data analysis?
2. **Explore→Explain** As scientists fear, it can **take extra time to learn to use high-end specialized tools** (e.g., Photoshop) favored by graphic artists and outreach professionals. But, if tools scientists want to use for exploring and analyzing their data produce publication or outreach-quality graphics by default, scientists will **save the time** they or a hired expert would need to spend optimizing a(n explanatory) graphic for distribution. The reduction in effort required to produce good graphics clearly improves scholarly communication amongst scientists, and the time savings has the potential to increase the amount of outreach material scientists generate as well.
3. **Explain→Explore** As the "democratization of data" (e.g., (12)) continues, more and more services are making data behind both scholarly journal figures and public outreach graphics freely accessible. These open data sets represent a wealth of new information that researchers can combine with more traditional data acquisitions in their inquiries. If it's **quick** and easy to get the data behind explanatory graphics, scientists will use those data, and learn more.

**How?**
Figure 2 offers a high-level view of how one new visualization environment, called "glue," developed by the authors, connects to several other data- and software-related services to enable a highly flexible system for facilitating easy travel between exploratory and explanatory data visualization.[3] The astronomical example in Figure 2 (see caption for details) shows a screenshot of several data sets being used in concert to understand star formation near the Orion nebula. But, a very similar figure could have been constructed using the high-dimensional medical and geographic information science (GIS) data sets to which the glue software is also extremely well-suited. Appendix 1 shows how Figure 2 would be altered were we to replace its astronomical focus with a medical or GIS example. Under the heading "Undisciplined Visualization," we explain below why sharing software amongst fields with similar kinds of data often makes sense, and sometimes does not.

---

[3] The glue software is described and documented on its website (13). Its functionality and flexibility via both a GUI and built-in command line is explained in (14), and its origins are explained in (15).

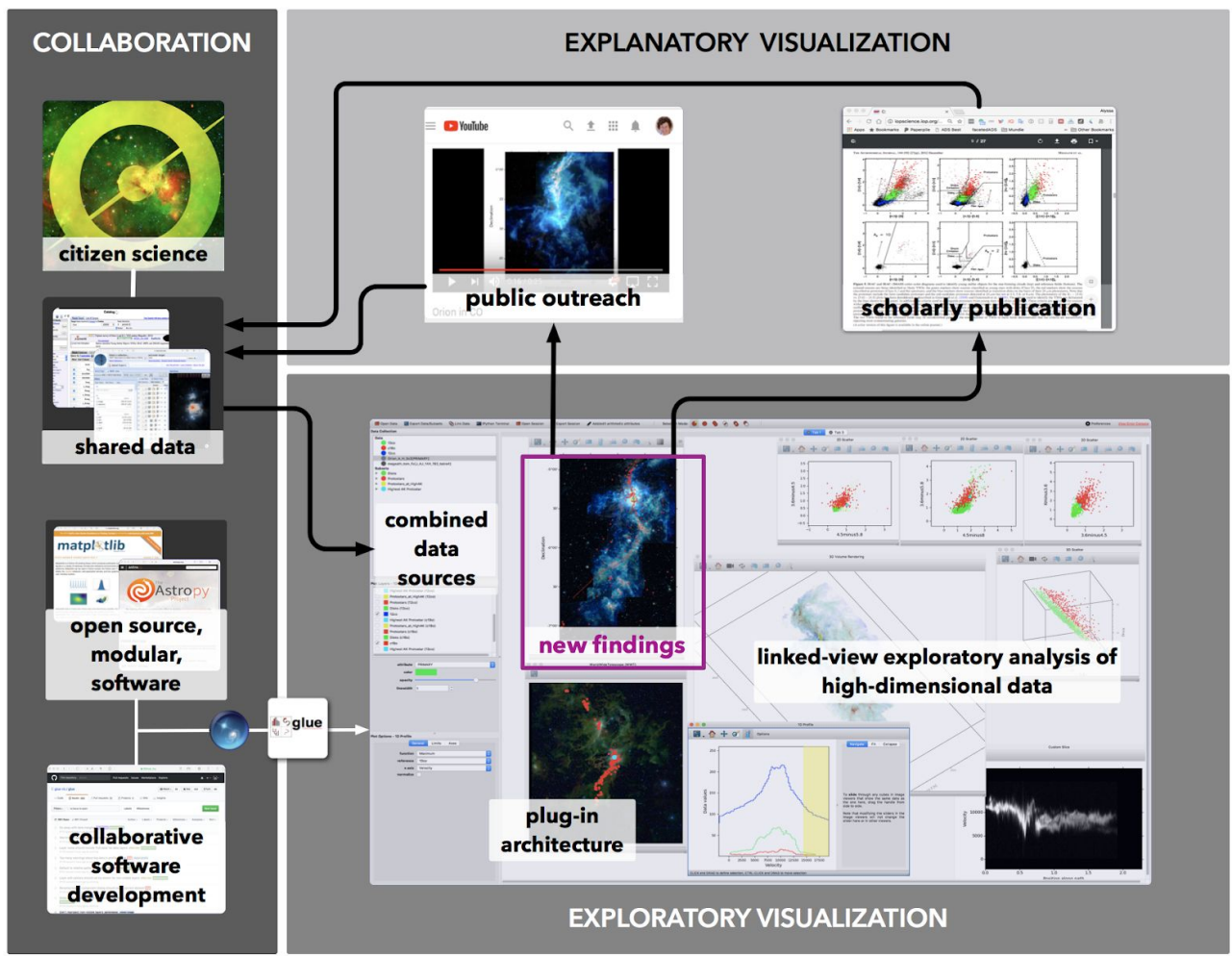

**Figure 2:** Connections and approaches making easy travel from Exploratory to Explanatory Visualization possible, in Astronomy, using glue. The **"combined data sources"** shown in the figure include the CARMA+NRO Survey of Carbon Monoxide (CO) gas in Orion (16) and the VISION Near-Infrared Survey of Orion (17). The **"scholarly publication"** (18) screenshot shown highlights statistical graphics (a set of *x-y* plots) describing the properties of young stars as measured by their color. The data from that publication (18) have been deposited by its author into the "Vizier" data repository shown in the **"shared data"** window, from which an astronomer can retrieve this information in order to add it to the "combined data sources" needed to recreate the published plots in glue. The user can also create new visualizations that combine the CO and infrared data (which together describe the three dimensional distribution and properties of the gas from which stars form) with the previously published catalog of young stars (which shows the positions and properties of young and forming stars). In the window labeled **"plug-in architecture,"** the young star catalog is superimposed on yet more contextual maps, available via the "WorldWide Telescope" (web-based) plug-in, which connects glue to dozens of additional freely-accessible maps of the Sky made as part of large collaborative projects, spanning nearly the whole observable electromagnetic spectrum. This particular screenshot from glue shows a user's **"new finding"** of bubbles caused by energetic winds from young stars permeating the gas (shown as a 3D rendering under the **"linked-view exploratory analysis of high-dimensional data"** label, and in several of the other open linked plots and image views as well). The panel labeled **"public outreach"** shows a YouTube video highlighting the new findings, which links to: the shared data, **"open-source, modular software,"** and **"collaborative software development"** platforms used in the work, and to the Zooniverse Milky Way Project **"citizen science"** site, where those interested can participate actively in research on star formation in the Milky Way Galaxy.

[ Note to the Editor: A video offering a tour of this figure, and/or a link to the YouTube video corresponding to the

> *"public outreach" panel, can be posted as supplements to this paper on the PNAS web site. We would also like to offer an interactive version of the figure, enabling panel-by-panel hyperlinks, zoom and/or description, either at the PNAS site or at a URL we can host. FYI, the figure was created using the "OmniGraffle" software, which offers easy export of html image maps. The figure in this document now is a relatively low-res preview of the zoomable, high-res, interactive, version we can provide.]*

One of the key aspects enabling the creation of customizable, highly flexible, connective, linked-view visualization systems like glue is the spirit of **collaboration** that characterizes the open-source software world, and much of the scientific world today. There are essentially four collaborative systems that enable the work highlighted in Figure 2. Moving from the bottom to the top of the figure, they are:

1. **Collaborative software development.** Some of world's best software is created under what many refer to as the "hero" model of development, where a lone genius writes thousands of lines of code. Even though others may be involved in the coding project, the hero represents a single-point failure mode. That model still works, but it's much better if the hero has friends, or if friends working together can outperform even a hero. The wildly-popular GitHub platform has now facilitated much easier "working together" options, by building a social interface to the version control git system introduced in 2005. Done well, collaborative software development can far outshine the hero model in terms of productivity, and, importantly, produce structured, modular, code that others can easily fork or re-use.

2. **Open-source, modular software**. Thanks in large part to easy-to-use collaborative platforms like GitHub, most developers today are happy to share code if their situation[4] permits. To enable re-use of their work, many developers, projects, and companies purposely design their software to be as modular as possible, much like the snap-circuit pieces shown in Figure 3, below.

   The glue software shown in Figure 2 is internally modular, and makes use of plug-in architecture to allow developers and users to build both data loaders and plot types that are useful in particular domain areas. In Figure 2, we show the WorldWide Telescope (WWT) web-based plug-in running inside of glue[5], built using WWT's JavaScript API. Other glue-related plug-ins include data loaders for medical and GIS image formats and plot types such as dendrogram (tree) and network diagrams. Individual users can configure glue to load only the modular plug-ins they may want to use.

   The brightest future for modularity likely now lies within web browsers. Flexible general-purpose visualization tools and languages like d3 and Vega are pervasive, and frameworks for deploying visualizations alongside analysis tools (e.g. machine learning using TensorFlow (19)), such as Jupyter Lab or Observable, make[6] excellent use of them. [See Jeff Heer's contribution to this volume for more on d3 and Vega.][7]

3. **Data sharing.** Owing to both carrot ("*hey, people, use my great data!*") and stick ("*no more government funding if you don't deposit your data*") approaches, data sharing is on the rise. Thus, tools that easily tap into a wide variety of sources at once have a competitive advantage. In the case of glue, this advantage is maximized by not only allowing users to load multiple datasets at once, but also allowing them to "glue" shared or mathematically-related attributes together, without the tedium and complexity of merging the wide variety of data types a user may want to bring to bear in any particular inquiry.
4. **Citizen science.** As John Tukey pointed out early on (10), and as most modern visualization researchers would agree (20), humans are better at some tasks than computers. Thanks to machine learning (e.g. (21)), the space where humans are needed is shrinking, but it may never fully disappear. In particular, humans are better at seeing patterns no one knew to look for, and interpreting them using domain knowledge. Easy flow back and forth between data repositories, citizen science tools, and visualization tools is facilitated by 1-3 above, and enables the kinds of unprecedented collaborative discoveries highlighted in the Citizen Science portion of this issue.[8]

---

[4] In academia, most code can be developed in the open. In industry, this is increasingly possible, as companies shift from a proprietary software business model to one where services around the software are profit centers.

[5] Technical note: glue presently runs on desktop machines (PC, Mac or UNIX), as a python program, using the qt windowing system.

[6] An alpha version of glue that runs entirely in the browser, within Jupyter Lab, shown by Goodman as a video at the Sackler Colloquium, will be released in late 2018. Note to editor--depending on timing, we'll want to include an actual link here.

[7] Note to editor--Assuming Heer is writing a chapter, we'll want to update this sentence with details.

**Always?**

The top two entries on the "10 Questions to Ask When Creating a Visualization" blog (22), intended as a guide for researchers and educators creating visualizations, are:

1. **Who** | Who is your audience? How expert will they be about the subject and/or display conventions?
2. **Explore-Explain** | Is your goal to explore, document, or explain your data or ideas, or a combination of these?

The answers to these two questions together determine how possible, or impossible, it is to create exploratory and/or explanatory visualizations suitable for *both* experts and the general public using a unified approach that relies only on a limited set of tools with which their user is comfortable. Sometimes, even though tools permit easy explanatory-exploratory travel, sociology or culture prohibits it. By way of a very simple example, consider color. To a physicist portraying temperature, the color blue encodes "hot," since bluer photons have higher energy, but in popular Western culture, blue is used to mean cold. So, a figure colored correctly for a physicist will not necessarily work for public outreach. Still, though, a physicist's figure produced in an exploratory system like the one portrayed in Figure 2 would work fine as an explanatory graphic for other physicists reading a scholarly report on the new findings.

**UNDISCIPLINED VISUALIZATION**

The glue software featured in Figure 2 is designed to be extraordinarily flexible and usable across disciplines. As we laid out in 2012 (15), our intent in creating glue was to extend the Exploratory Data Analysis principles of Tukey (10) beyond tabular (ASCII) data (e.g., DataDesk, Spotfire) and maps (e.g., ArcGIS), or even combinations of those (e.g., Tableau). Our goal was to move Exploratory Data Analysis[9] into a higher-dimensional world where the kinds of real-time interactions (selections, manipulations) with data Tukey envisioned in the 1970s would be linked across all open statistical graphs, tables, diagrams, 2D images, or 3D volumetric displays. In glue, if a researcher wants to interactively or algorithmically select a tumor in a brain scan, or a star-forming region in a molecular cloud, or a cluster of errant airplanes, she can perform any of these actions just as easily as she can select a cluster of points in Tableau or Spotfire.

---

[8] Note to editor--we'll want to link to the appropriate articles here.
[9] The techniques, concepts, and terminology for "direct manipulation" in linked-view systems were developed beyond what Tukey envisioned by pioneers like Paul Velleman (23), whose work generated DataDesk in the 1980s, Ben Shneiderman, whose work led to Spotfire (24) in the 1990s, and by Stolte, Tang and Hanrahan whose paper on "Polaris"(25) ultimately led to Tableau in 2003.

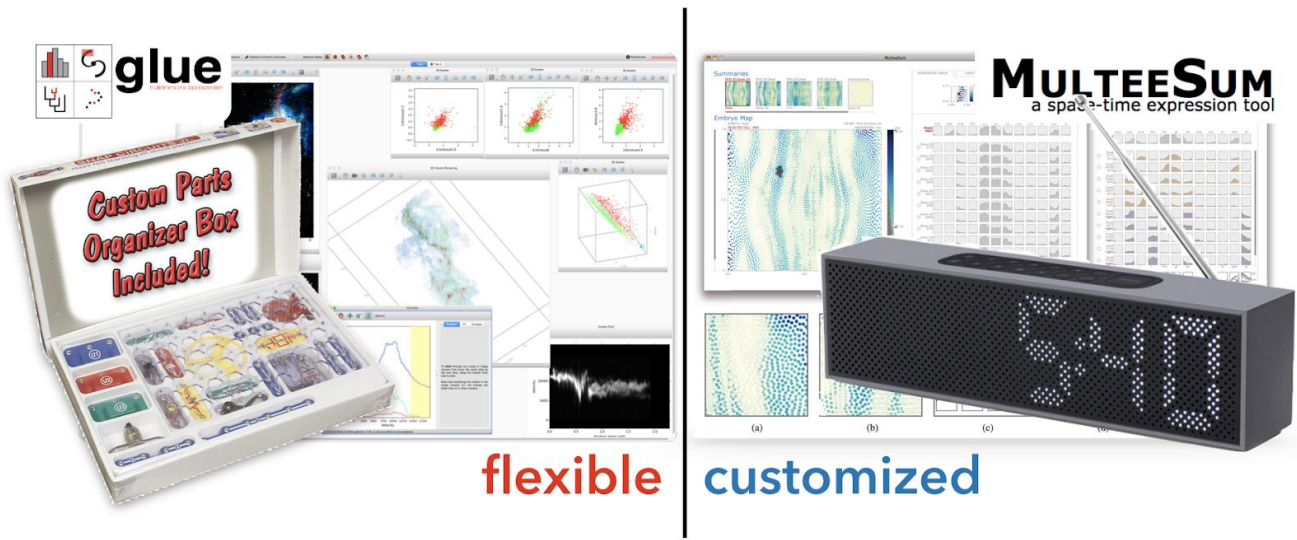

**Figure 3:** Flexible and customized electronic circuits, and analogous visualization environments. Pictured as flexible exemplars are a "Snap Circuits" kit used in building a wide variety of electronic devices along with the glue software discussed in this paper. On the customized side, we show the beautiful MulteeSum software (26) and an elegant, special-purpose, clock radio.

Flexibility in a visualization system has advantages and disadvantages. We discuss below, in turn, two trade-offs around flexibility: the first concerns software architecture, and the second, harder-to-judge, issue concerns disciplinary scientists' capacity to generalize when it comes to high-dimensional data.

**Flexible vs. Custom Software**

The flexibility tradeoff around visualization software is represented by analogy in Figure 3. If a person wants a clock radio, her best bet is to buy an elegant device designed specifically to display time and play audio. On the other hand, if she does not quite know what she wants, and that may or may not include a clock, or a radio, then a well-designed kit, like the "Snap Circuits" one shown in Figure 3, would be a better bet.

When funding and interest permit, fantastic, special-purpose, custom software tools can be designed by visualization experts working together with scientists. Carefully planned design studies (20) supported by relevant research, produce these tools, and careful evaluation shows that they hold the potential to dramatically optimize efficiency and discovery (e.g., 27).

The MulteeSum tool (26) used as an exemplar in Figure 3 supports "inspection and curation of data sets showing gene expression over time, in conjunction with the spatial location of the cells where the genes are expressed," and it allows its user to load and interrelate several datasets at once. Given the similarity of MulteeSum and most glue users' goals (linking high-dimensional datasets, and views of them, together in a real-time visualization analysis environment), one could, in principle, implement most or all of MulteeSum's functionality in glue, adding plug-ins where needed. But *should* anyone do that? MulteeSum was presented in 2010, long before glue had the capabilities it does now, so this question would have been nonsensical until recently. The right answer here lies in a re-thinking of the level of modularity for visualization software, as is considered[10] in this volume by Jeff Heer and Maneesh Agrawala.

Imagine building software like MulteeSum using not relatively low-level graphics languages like d3 (or processing), but instead higher-level, more semantically-meaningful, pieces, like those enabled by Vega, inside of an environment that looks like the Jupyter Lab version of glue discussed above. In such an impending world, not only will the line separating exploratory and explanatory visualization (Figure 2) blur, but also the line between flexible and custom tools (Figure 3).

**Data-Dimensions-Display**
Through training and experience, scientists learn to think about particular kinds of plots or data displays common in their own discipline as having particular meanings or uses. As a result, it's not always easy for disciplinary scientists to "undiscipline" themselves when using a super-flexible visualization system that is agnostic about which discipline's data is being displayed. The cognitive load associated with generalizing beyond what is familiar is not to be underestimated.

When showing new users their own data in glue, which is arguably the most flexible high-dimensional linked-view visualization environment available today, one often hears comments like *"wow, people don't usually plot it that way… but it's really cool!,"* or, *"that looks amazing, but what does it mean?,"* or *"wait, my head's going to explode, what's that?"* We use the term **"newness overload"** here to summarize these cognitive effects.

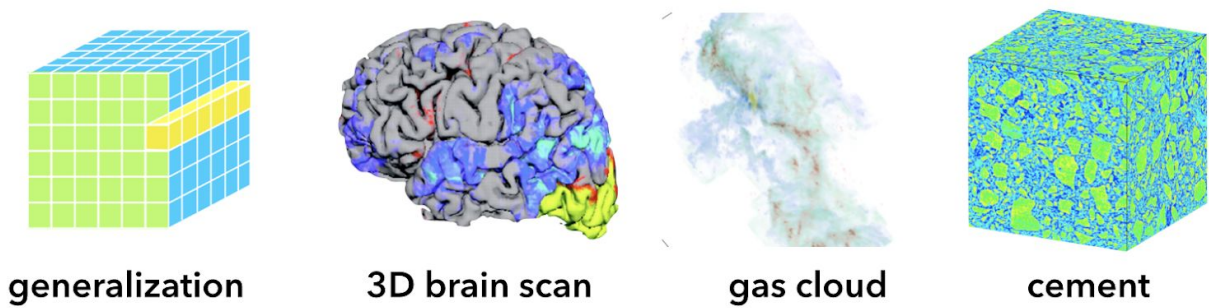

**Figure 4:** High-dimensional data spaces. From left to right, a generalized representation of 3D voxels, a brain scan shown as a surface rendering, the Orion cloud shown in glue from Figure 2, and section of a block of cement.

When explaining to scientists why doctors and astronomers should work together (28, 29), we often show a set of images like the ones in Figure 4. The left-most panel of the figure offers a generic representation of a 3D cube of voxels, as well as a 1D column of voxels (shown in yellow) and a sample 2D plane of voxels (shown in green). This simple representation, irrespective of whether the dimensions are spatial or not, proves useful in "undisciplining" researchers too comfortable in their own domain.

---

[10] Note to editor--without seeing all the contributions at once, we cannot finalize this paragraph. We are happy to update it to reference other chapters in the refereeing/editing process.

Medical and astronomical researchers pretty much all understand the word "image" to mean a 2D representation of intensity, but only some appreciate that both astronomers and radiologists build 3D volumetric representations, like those shown in Figure 4, by combining many 2D "image" planes of data at once.  Worse, in each discipline,  particular planes, or sums of planes, are given completely different jargon-laden names.  For example, an "axial view" in brain imaging (according to Wikipedia) is a "radiographic projection devised to obtain direct visualization of the base of the skull," and a "integrated intensity map" in astronomy sums all emission along the observer's line of sight. In practice, these two dramatically-differently-named images are exactly the same from a mathematical point of view–just a sum (a.k.a. "projection") of all the image planes along one direction.  The same sum could be made for the block of cement in Figure 4, and it would no doubt have a third jargon-laden name.

In principle, researchers seeking to visually explore high-dimensional data in any discipline should always think deeply about the connections between the inherent dimensionality of their data, and the number of dimensions they can effectively display at once.  We refer to that kind of generalized view of optimizing visualizations  as "data-dimensions-display," and the word "dimensions" need not refer to orthogonal or even spatial dimensions.  But, given the newness overload[11] offered to disciplinary scientists by glue's flexibility, training glue users to think in this "undisciplined" way is challenging.  We discuss this challenge in our concluding thoughts about the future, below.

**THE FUTURE: TRAINING, CONVERSATION, COLLABORATION**

**Training**

To realize the full discovery potential of merged exploratory and explanatory visualization approaches, we need to train scientists and visualization researchers to understand each other's worlds more deeply.  The 2006 NIH-NSF Visualization Research Challenges report (6) came to much the same conclusion in, so what has changed?  Urgency.  In 2006, no one quite knew what a "data scientist" was, but today, those words describe one of the most in-demand, high-paying, professions of the 21st century.  Data volume is rising faster and faster, as is the diversity of data sets available – both in the commercial and academic sectors.  Despite the rise of data science, though, today's students are typically not trained–at any level of their education–in data visualization.  Even the best graduate students in science at Harvard typically arrive completely naive about what visualization researchers have learned about how humans perceive graphical displays of information.

Over the past decade or so, more and more PhD students in science fields are taking computer science and data science courses.  These courses often focus almost entirely on purely statistical approaches to data analysis, and they foster the idea that machine learning and AI are all that is needed for insight.  They do not foster the ideas that one of the 20th century's greatest statisticians, John Tukey, put forward (8), about visualization: 1) having the potential to give unanticipated insight to later be followed up with quantitative, statistical, analysis; or 2) that algorithms can make errors easily discovered and understood with visualization.

---

[11] By way of specific example, if one zooms in carefully on the panel labelled "New Findings" in Figure 2, one will notice a squiggly red line.  That red line is a user-defined non-orthogonal dimension allowed in glue, and the grey scale plot in the extreme lower-right corner of Figure 2 is essentially an image created by summing the emission in the CO cube shown in the volumetric rendering onto a ribbon whose projection is the red line, and whose width is that of the full cube. Expert radio astronomers call this a "position-velocity" or "p-v" diagram *when a cut like the one shown in red is made along a straight line.* Standard astronomy software does not allow cutting along a non-orthogonal, user-defined dimension, so there is no name for this new kind of "*p-v*" image, and it takes several minutes to explain its value, even to an expert.

It seems time for today's scientists to develop visualization skills alongside the math and computer science skills they already need. Some of this visualization training will be relatively easy, and will narrow the gap between exploratory and explanatory visualization. For example, nearly all visualization researchers know that certain color tables emphasize certain aspects of a data set (e.g., 30), like the one shown in the top row of Figure 5. It is the rare scientist, though, who considers color choices as carefully a visualization researcher would, or who thinks about the potential deception created by the mathematical transformations offered by buttons in a program like glue labeled "linear," "square root," "arcsinh," and "log." Instead, scientists and outreach professionals choose what is aesthetically pleasing to them, and as a result both scholarly journals and media are filled with images that may not quite have the meaning scientists intended. Published images also often hide meaning that could have been revealed had the image creator made a more informed choice.

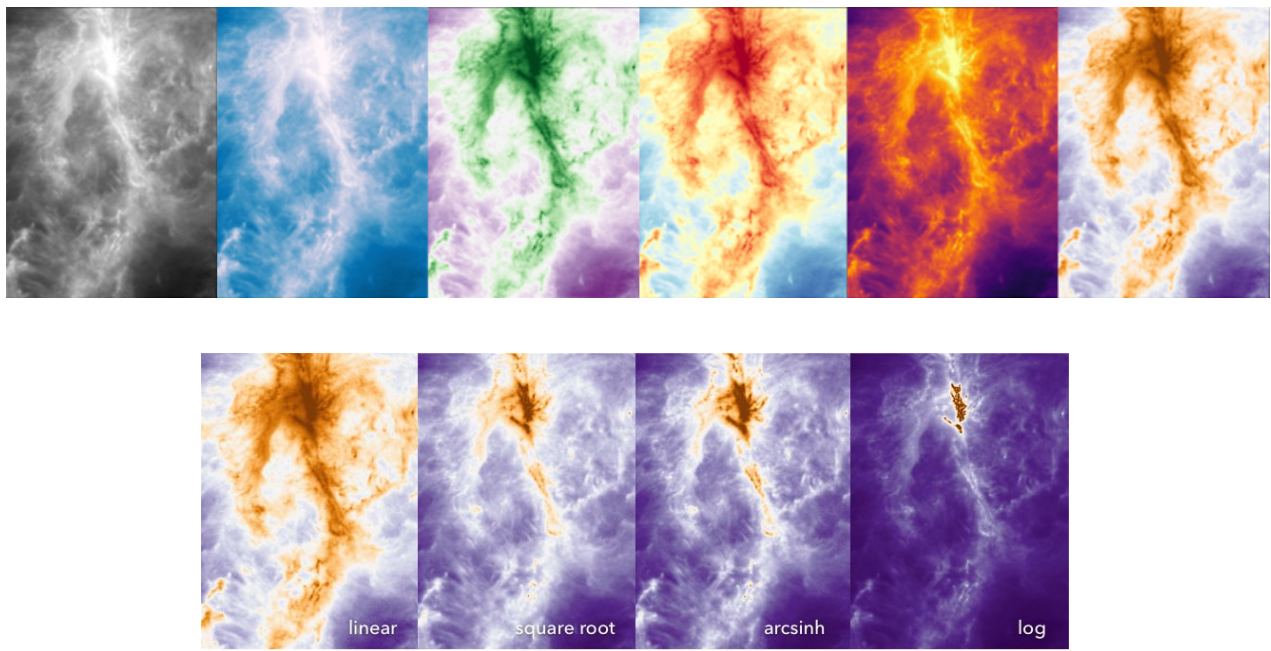

**Figure 5:** Top row: six different color tables (all with linear stretch) applied to exactly the same data set (Orion as seen by the Herschel Space Observatory). Bottom row: four different color stretches applied to the image at the right-most edge of the top row, as labeled in the figure.

Another easy win without any new research will come from scientists' adoption of existing technologies for publishing online interactive data graphics, as demonstrated in the 2015 *"Paper" of the Future* mockup commissioned by the American Astronomical Society (31).

**Concluding thoughts on Conversation and Collaboration**
As we stated at the outset, "*Access* to software and *proficiency* at using it to gain or communicate insight are not the same thing." Improving training is clearly likely to improve proficiency, but it will not on its own create new research at the science-visualization boundary that is needed to accelerate progress in both fields. Training will allow for "new thinking with" visualization, but to get to "new thinking on" visualization, we need new conversations and collaborations.

To spark the needed conversations, we conclude with two provocations, and one suggestion for collaboration.

*Provocation 1*: Computer scientists need to stop using the word "workflow" to describe scientists' process. As Figure 2 hopefully indicates, a scientist can start at any point in the explanatory or exploratory visualization regimes, or from shared data, to pursue an investigation. The process can look random, and iterative, to an outsider. It is fine to think of small pieces of the full process as "workflows" or "tasks," but computer scientists importantly need to understand that scientists do not think about what they do as a flow from one task to another--they would never use those particular words, so some kind of new descriptors and vocabulary, more suitable to how science research really proceeds, are needed. In this paper, we have tried to explain that there is no longer even a one-way "workflow" from exploratory to explanatory work, so it would be beneficial to look for some new, more mutually understandable, language.

*Provocation 2:* Scientists have to invest the time to understand visualization research, and opportunities to train themselves and their students in effective visualization techniques. Just as learning new statistical techniques is absolutely critical in today's "data science" world, so are visualization techniques, and the two are critically related.

*Suggestion:* As co-authors, we represent a professor in science (Goodman), a professor in computer and information science (Borkin), and an expert developer trained in both science and computer science (Robitaille). The conversations the three of us have had over the course of our collaboration, and especially in creating this paper, have made it clear to us that many more purposefully "translational" papers like this one, along with similarly-intentioned videos and conferences like the Creativity and Collaboration Sackler Colloquium, are needed to facilitate breakthroughs made by scientists and visualization experts working together to advance not only each other's fields but also the fertile one that lies in-between.

## ACKNOWLEDGEMENTS

We thank Ben Shneiderman for imagining and realizing the Sackler Colloquium that precipitated this work. Sackler co-organizer Maneesh Agrawala, and speaker Curtis Wong's excellent insights also contributed to the views we offer here. Figure 2 benefited greatly from discussions with Felice Frankel, Ben Shneiderman and Catherine Zucker. The CARMA-NRO Orion Survey (16) and VISION Survey in Orion (17) teams generously contributed their data and expertise to the glue example highlighted in Figure 2. Christopher Beaumont, co-founder and developer extraordinaire on the glue project, also has our thanks, as does graduate student/glue developer Penny Qian.

*Submitted as an invited "Perspectives" Paper for PNAS, in conjunction with the 2018 Sackler Colloquium*

**Appendix (or Table) 1 – Collaboration, Exploration, and Explanation: Examples across Three Disciplines[12]**

| | ASTRONOMY | MEDICINE | GEOSPATIAL SCIENCE |
|---|---|---|---|
| graphic useful in *explanation or exploration*, produced *for outreach or for publication* | 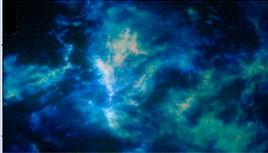 Orion public outreach image[13] | 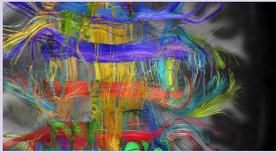 fundraising PR brain pathways image[14] | 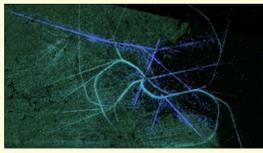 satellite image with 3D plane paths |
| shared data | e.g., MAST or Vizier | e.g., NIH repositories | Flightradar24 API, Google Maps (API) |
| *discipline-specific* open source, modular software | astropy | pydicom, DiPy, NiBabel | rasterio, pyModeS & adsb-sdr |
| "undisciplined" open source, modular software | Qt, Matplotlib, Jupyter Lab | | |
| collaborative software development | github | | |
| linked-view exploratory analysis of (combinations of) high-dimensional data (sets) | glue | | |
| plug-in options | WorldWide Telescope | tree/network visualizer | geospatial file loader |
| new findings | evacuated cavities surrounded by gas | connections, disease, abnormalities | take off & Landing trajectory differences |
| Export options for outreach/publication | new image, web interactive or video, or 3DPDF | | |

---

[12] Note to editors: Websites and software are hyperlinked in this table for now. If PNAS wants citations done another way, we can make a change later. We also do not know, for Perspectives, whether this is more suitable as a Table in the main text or as and "Appendix" or "Supplementary Information."
[13] YouTube citation will be updated once video is released to the public by NSF (June 2018).
[14] From https://giving.massgeneral.org/mass-general-brain-mapping-reveals-pathways-human-thought/